\documentclass[prx,reprint,amsmath,amssymb,amsfonts,twocolumn,nofootinbib,showpacs]{revtex4}

\usepackage{bm,graphicx,mathrsfs}
\usepackage{graphicx}
\usepackage{epsfig}
\usepackage{amsmath,bbm}
\usepackage{amsfonts,amssymb}
\usepackage{times}
\usepackage{dsfont}

\newcommand{\be}{\begin{equation*}}
\newcommand{\ee}{\end{equation*}}
\newcommand{\bea}{\begin{eqnarray}}
\newcommand{\eea}{\end{eqnarray}}

\newcommand{\ket}[1]{|#1\rangle}
\newcommand{\bra}[1]{\langle#1|}

\def\>{\rangle}
\def\<{\langle}

\begin{document}

\title{Steady state entanglement of two superconducting qubits engineered by dissipation}

\author{Florentin Reiter$^{1}$\footnote{reiter@nbi.dk}, L. Tornberg$^{2}$, G\"{o}ran Johansson$^{2}$ and Anders S. S\o rensen$^{1}$}
\affiliation{${ }^{1}$QUANTOP, Niels Bohr Institute, University of Copenhagen, Blegdamsvej 17, DK-2100 Copenhagen, Denmark\\
${ }^{2}$Chalmers University of Technology, SE-41296 Gothenburg, Sweden}
\date{\today}

\begin{abstract}
We present a scheme for the dissipative preparation of an entangled steady state of two superconducting qubits in a circuit QED setup. Combining resonator photon loss, a dissipative process already present in the setup, with an effective two-photon microwave drive, we engineer an effective decay mechanism which prepares a maximally entangled state of the two qubits. This state is then maintained as the steady state of the driven, dissipative evolution. The performance of the dissipative state preparation protocol is studied analytically and verified numerically. In view of the experimental implementation of the presented scheme we investigate the effects of potential experimental imperfections and show that our scheme is robust to small deviations in the parameters. We find that high fidelities with the target state can be achieved both with state-of-the-art 3D, as well as with the more commonly used 2D transmons. The promising results of our study thus open a route for the demonstration of a highly entangled steady state in circuit QED.
\end{abstract}


\maketitle

\section{Introduction}

One of the most peculiar properties a physical system can exhibit is quantum-mechanical entanglement \cite{ES}. From a fundamental perspective, entanglement is a non-classical effect which is indispensable for the understanding of fundamental quantum physics. From a technological perspective, entanglement is useful for enhanced measurement techniques and is an important element in quantum information processing and quantum communication \cite{NC}. For the past two decades great effort has been invested into the generation and investigation of entangled states. Inspired by the circuit model of quantum computation, entanglement has predominantly been investigated by means of coherent interactions, i.e. by applying sequences of unitary gates.
Today, there is a large number of physical systems where entanglement has been demonstrated and which are considered suitable for the realization of advanced quantum information protocols.
Out of these, superconducting systems \cite{RevSC} have proven to be good candidates for the realization of quantum algorithms involving many gate operations \cite{Neeley, Fedorov, Reed}. Despite impressive reductions of the decoherence in superconducting systems \cite{Houck, Chow, Rigetti, Poletto, Paik, Sears}, any state other than the ground state will deteriorate over time. As a consequence, today's quantum computation and simulation are still limited to elementary protocols on small scales.


Over the past few years, however, an alternative approach of dissipative state engineering, dissipative quantum computing and dissipative phase transitions \cite{VWC, Kraus, Diehl} has emerged and gained increasing attention. As opposed to unitary quantum computing, where decoherence and dissipation act detrimentally on the state preparation process and on the prepared state, the central idea here is to prepare non-trivial quantum states relevant for quantum information, simulation \cite{Diehl, Muller}, memories \cite{DissMemory}, or communication \cite{DissRepeater} by means of an engineered interaction of the system with its environment.
As opposed to unitary methods, dissipative quantum computation and dissipative state engineering involve steady states. Such states are resilient to the dissipative evolution by which they have been produced. This provides an additional stabilization against other kinds of decoherence.
The question of whether this new dissipative paradigm can become an alternative or even superior approach to unitary quantum information processing can, however, not be answered in a single step. Instead, exploration of its capabilities needs to begin at a small scale. Here, an elementary quantum information processing task is found in the preparation of a maximally entangled Bell state as the steady state.

Previous theoretical work on entanglement generation utilizing dissipation has dealt with a number of quantum optical and solid state systems, in particular cavity QED \cite{PHBK, Clark, VB, WS, RKS, Busch, KRS}, atomic ensembles \cite{Parkins, MPC, DallaTorre}, ion traps \cite{PCZ, CBK, Muller, Barreiro}, plasmonic systems \cite{AGZ, Gullans, GP}, light fields \cite{Kiffner} and optical lattices \cite{Diehl, FossFeig}. The first experimental demonstrations were achieved in atomic ensembles \cite{Krauter} and ion traps \cite{Barreiro}. Several different state preparation tasks involving dissipation have also been considered for superconducting systems \cite{Zhang, Li, Xia, Murch}. So far, generation of maximally entangled steady states in the widely used setting of two superconducting qubits coupled through a common resonator has not been studied. In this work, we consider the dissipative preparation of a maximally entangled state in this system.

As opposed to previous studies of atomic systems coupled through a common resonator \cite{KRS,RKS} the realization of similar effects in superconducting systems raises a number of additional challenges. These are (1) a different energy level diagram, (2) additional, undesired transitions between qubit levels since these are not, as in atomic systems, suppressed by selection rules, and (3) additional decoherence mechanisms acting on the qubit. In addition, the dissipative entangling operation shall be independent of the initial state and reach a highly entangled steady state within reasonable time, also in the presence of imperfections in the setup.
We will, in the following, discuss a scheme for superconducting qubits which fulfills these requirements, surmounting the above challenges.

As detailed in Sec. \ref{SecSetup}, our scheme is specifically designed to exploit (1) the level structure of typical transmon qubits \cite{Koch}, which constitute weakly anharmonic oscillators. The scheme is, however, not particularly restricted to transmons, but can also be applied to phase qubits \cite{PhaseQubit} coupled to a resonator. Utilizing a coherent two-photon drive of a dipole-forbidden transition with a two-tone microwave field similar to Refs. \cite{Kelly, Poletto}, we engineer an effective resonator loss process which deterministically prepares the maximally entangled singlet state $\ket{\rm S}$, as is described in Sec. \ref{SecMechanism}. Here we also show that (2) the coupling of the resonator to several transitions of the transmon is in fact an advantage, as it provides a transfer from the undesired states to the one from which the target state $\ket{\rm S}$ is prepared. Given that $\ket{\rm S}$ is produced by a time-independent loss process and continuous wave fields, it is a steady state of the dissipative evolution.

In Sec. \ref{SecParameter}, we investigate the performance of our scheme, both analytically, to derive benchmarks for the protocol, and numerically, to verify the mechanisms that underlie the presented dissipative state preparation scheme. Our results show that a maximally entangled state of two superconducting qubits can be prepared rapidly and with a high fidelity, even in the presence of (3) realistic qubit decoherence rates and imperfections. High fidelities are obtained both for state-of-the-art 3D, as well as for the more common 2D transmons. By fulfilling the above requirements our proposal thus opens a route for the dissipative preparation of maximally entangled states of superconducting systems using existing technology.

\begin{figure}[t]
\centering
\includegraphics[width=\columnwidth]{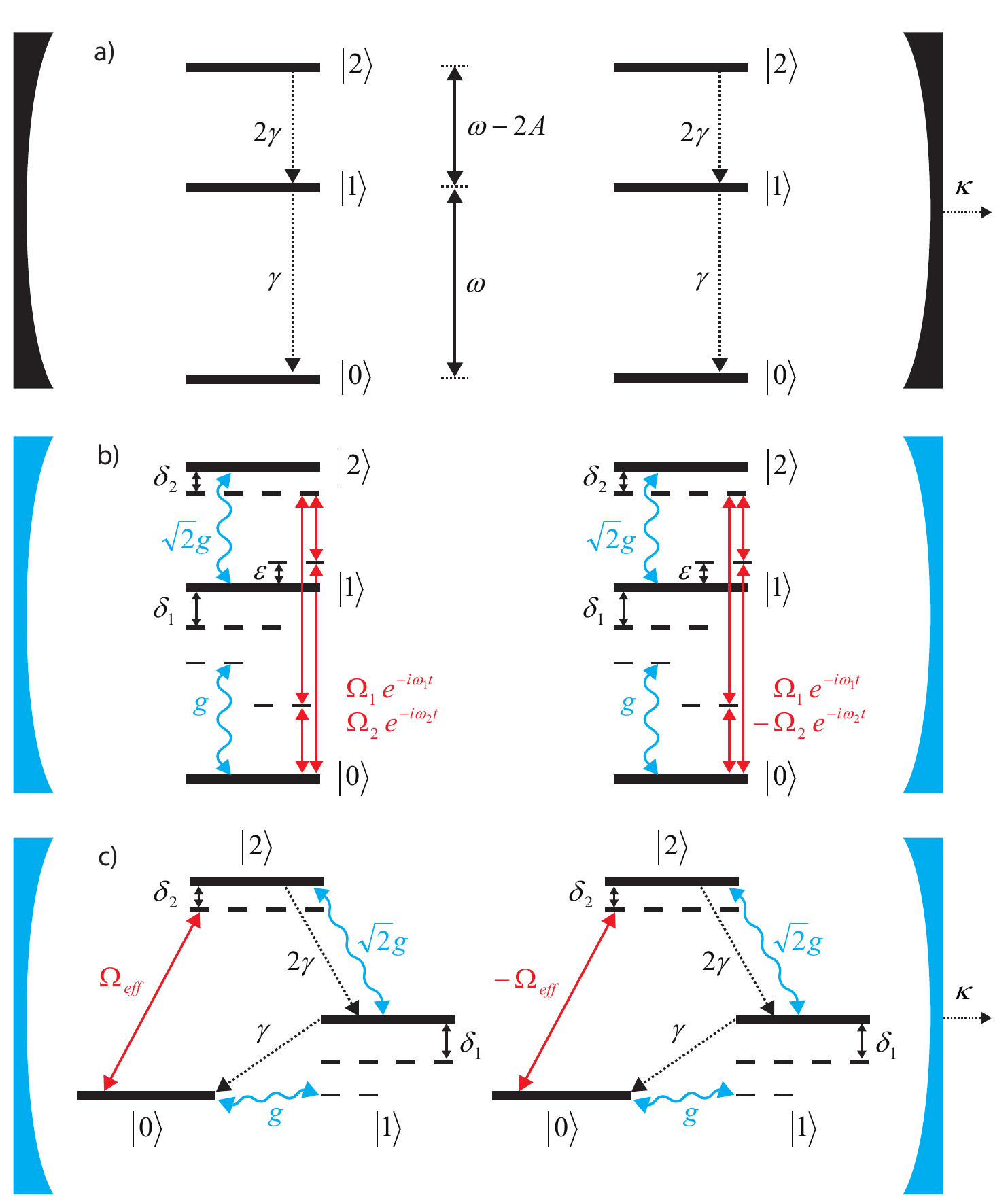}
\caption{(Color) Setup. The internal levels of two transmons (a) are coupled by coherent interactions (b) to mimic the $\Lambda$ system in (c). Two microwave fields $\Omega_{1/2}$ provide virtual couplings of the transitions $\ket{0} \leftrightarrow \ket{1}$ and $\ket{1} \leftrightarrow \ket{2}$ (b) which combine to an effective two-photon drive $\Omega_{\rm eff}$ of the transition $\ket{0} \leftrightarrow \ket{2}$. The transmon-resonator coupling ($g$) is resonant with the upper transition and detuned by $\delta_1-\delta_{\rm c}$ from the lower transition. Spontaneous emission ($\gamma$) and resonator photon loss ($\kappa$) are present as decoherence processes. The detunings are defined in the text.}
\label{FigSetup}
\end{figure}

\begin{figure*}[t]
\centering
\includegraphics[width=17.2cm]{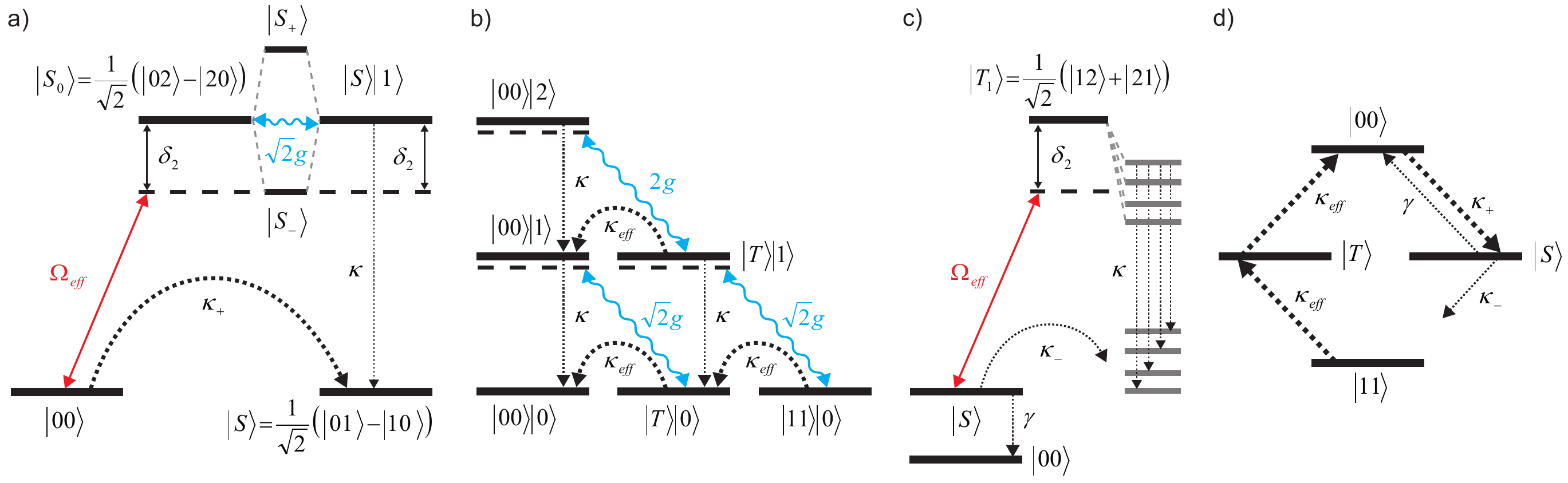}
\caption{(Color). (a)-(b) Dissipative state preparation mechanisms, (c) loss mechanisms, and (d) effective lower-level decay processes. (a) Effective resonator decay from $\ket{00}$ into $\ket{\rm S}$ involves coherent coupling to $\ket{\rm S_0}$. $\ket{\rm S_0}$ and $\ket{\rm S}\ket{1}$ are strongly coupled ($\sqrt{2} g$) so that these states hybridize and form dressed states $\ket{\rm S_\pm}$ (shown here for a choice of $\delta_{\rm c} = \delta_2 - \delta_1$, by which the resonator is resonant with the upper transition). By setting $\delta_2=\sqrt{2}g$ the driving from $\ket{00}$ is resonant with the lower dressed state $\ket{\rm S_-}$. Population from $\ket{00}$ is thus rapidly excited and decays into $\ket{\rm S}$ via the effective engineered resonator decay $\kappa_+$. (b) The population of the bright states $\ket{11}$ and $\ket{\rm T}$ is shuffled to $\ket{00}\ket{0}$ by the resonator coupling $g$ and successive resonator decay at an effective rate of $\kappa_{\rm eff}$. (c) The two-photon drive also causes an undesired coupling of the otherwise dark target state $\ket{\rm S}$ to an excited state $\ket{\rm T_1}$. $\ket{\rm T_1}$, in turn, couples to a number of (resonator-) excited states which form dressed states at different energies (indicated) and eventually decay to other states. These can generally be made off-resonant with the drive from $\ket{\rm S}$ by an appropriate choice of the resonator and microwave detunings so that the effective resonator decay $\kappa_-$ from $\ket{\rm S}$ is suppressed. In addition, since $\ket{\rm S}$ is a dark state of the cavity interaction, the only direct decay mechanism is through the weak qubit decay $\gamma$ to $\ket{00}$. The effective decay processes of the lower levels are summarized in (d).}
\label{FigMechanism}
\end{figure*}

\section{Setup: coherent and dissipative interactions of two coupled transmons}
\label{SecSetup}

For our study we consider two superconducting transmons \cite{Koch} coupled to a common resonator in a circuit QED setup. The coherent dynamics of the system is described by a Hamiltonian $H = H_{\rm free} + H_{\rm cav} + H_{\rm d}$. The energy levels are illustrated in Fig. \ref{FigSetup} a) and described by the free Hamiltonian
\begin{align}
H_{\rm free} = &\omega_{\rm c} a^\dagger a + \sum_{j=1,2} \left(2 \omega - 2 A\right) \ket{2}_j \bra{2} + \omega \ket{1}_j \bra{1},
\end{align}
with levels $\ket{k}$ of transmon $j$ and the resonator mode $a$. Here, $\omega$ denotes the level spacing of the two lower levels and $A$ the anharmonicity, with $\hbar = 1$. In our analytical discussion we will focus on the first three levels of the transmons, $\ket{0}$, $\ket{1}$ and $\ket{2}$. Our numerical assessment will also include the fourth level, $\ket{3}$. 

The transitions of the transmons, $\ket{0} \leftrightarrow \ket{1}$ and $\ket{1} \leftrightarrow \ket{2}$, are coupled by the coherent interactions shown in Fig. \ref{FigSetup} b). They are described by a Hamiltonian $H_{\rm cav} + H_{\rm d}$. Here, $H_{\rm cav}$ represents the coupling of the resonator to the transitions of the transmons,
\begin{align}
H_{\rm cav} = &\sum_{j=1,2} g a^\dagger \left(\ket{0}_j \bra{1} + \sqrt{2} \ket{1}_j \bra{2} \right) + H.c.,
\end{align}
with a coupling constant $g$, and a factor of $\sqrt{2}$ for the matrix element of the upper transition. The coherent drive
\begin{align}
H_{\rm d} = &\sum_{j=1,2} \left(\frac{\Omega_1}{2} e^{- i \omega_1 t} + (-1)^j \frac{\Omega_2}{2} e^{- i \omega_2 t} \right) \nonumber \times \\ &\times \left(\ket{1}_j \bra{0} + \sqrt{2} \ket{2}_j \bra{1} \right) + H.c.
\end{align}
contains several microwave fields which couple the transitions $\ket{0} \leftrightarrow \ket{1}$ and $\ket{1} \leftrightarrow \ket{2}$. We assume that the drive with $\Omega_1$ exhibits an identical phase, whereas the phase of $\Omega_2$ is opposite for the two transmons. This can be achieved by driving the qubits with the field $\Omega_1$ through a common wire and with the field $\Omega_2$ through additional individual wires, similar to Refs. \cite{Groot, Chow11, Filipp}. As we will see, this choice of phases allows us to break the symmetry of the system and thereby drive certain transitions which play an important role in our proposal.

We choose the frequencies of the two fields in such a way that they combine to an effective two-photon drive of the transition $\ket{0} \leftrightarrow \ket{2}$ with a coupling constant of $\Omega_{\rm eff}$ that will be derived in Sec. \ref{SecTwo}. In doing so, we render the couplings of the system resembling the $\Lambda$ system shown in Fig. \ref{FigSetup} c), with (meta-) stable lower levels $\ket{0}$ and $\ket{1}$ and an ``excited'' level $\ket{2}$ for each of the transmons. ``Excitation'' from $\ket{0}$ to $\ket{2}$ is then accomplished by the two-photon drive with $\Omega_{\rm eff}$. For most of this paper, we will assume that the resonator coupling is resonant with the transition $\ket{1} \leftrightarrow \ket{2}$, while being somewhat detuned from the lower transition $\ket{0} \leftrightarrow \ket{1}$.

In the following, we will avoid the fast dynamics in the drive by changing into a frame rotating with a Hamiltonian
\begin{align}
H_{\rm rot} = \bar{\omega} \left(a^\dagger a + \sum_k \sum_{j=1,2} k \ket{k}_j \bra{k}\right),
\end{align}
where $\bar{\omega} \equiv \frac{1}{2} \left(\omega_1 + \omega_2\right)$ is the mean frequency of the classical driving fields. Applying a unitary $\mathcal{U} = {\rm exp}[i H_{\rm rot} t]$ we obtain a transformed Hamiltonian $H^{'} = \mathcal{U} H \mathcal{U}^\dagger + i \dot{\mathcal{U}} \mathcal{U}^\dagger = H^{'}_{\rm free} + H^{'}_{\rm cav} + H^{'}_{\rm d}$ in a frame rotating with $H_{\rm rot}$. The transformed free Hamiltonian can be expressed as
\begin{align}
H^{'}_{\rm free} = \delta_{\rm c} a^{\dagger} a + \sum_{j=1,2} \delta_1 \ket{1}_j \bra{1} + \delta_2 \ket{2}_j \bra{2},
\end{align}
where $\delta_1 = \omega - \bar{\omega}$, $\delta_2 = 2(\omega - \bar{\omega}) - 2 A$, and $\delta_{\rm c} \equiv \omega_{\rm c} - \bar{\omega}$ denote the energies of the transmons and the resonator in the rotating frame. Furthermore, we obtain the interaction Hamiltonians $H^{'}_{\rm cav} = H^{}_{\rm cav}$
for the transmon-resonator coupling and
\begin{align}
H^{'}_{\rm d} = &\sum_{j=1,2} \left(\frac{\Omega_1}{2} e^{i \Delta_1 t} + (-1)^j \frac{\Omega_2}{2} e^{i \Delta_2 t} \right) \times \nonumber \\ &\times \left(\ket{1}_j\bra{0} + \sqrt{2} \ket{2}_j\bra{1}\right) + H. c.
\end{align}
for the drive. With this choice of the reference frame rotating with the mean frequency, we find the detunings of the microwave fields $\Delta_{1/2} \equiv \bar{\omega} - \omega_{1/2} = \pm \frac{1}{2}\left(\omega_2 - \omega_1\right)$.\\

In addition to the coherent dynamics discussed so far, the system also exhibits dissipative couplings, which is essential for the dissipative state preparation mechanisms we would like to engineer. The dissipative dynamics of the open system is determined by its coupling to the bath and the properties of the bath. Assuming the bath to be Markovian, the system dynamics is governed by a master equation of Lindblad form
\begin{align}
\label{EqMaster}
\dot{\rho}=i\left[\rho,H\right]+\sum_k L_k \rho L^\dagger_k - \frac{1}{2}\left(L^\dagger_k L_k\rho+\rho L^\dagger_k L_k\right),
\end{align}
with one Lindblad operator $L_k$ for each physical decay process present in the system. As illustrated in Fig. \ref{FigSetup} a), we assume that transmon $j$ undergoes spontaneous decay which in the transmon regime can be described by
\begin{align}
\label{EqSpont1}
L_{\gamma1,j} &= \sqrt{\gamma} \ket{0}_j\bra{1} \\
L_{\gamma2,j} &= \sqrt{2\gamma} \ket{1}_j\bra{2}.
\label{EqSpont2}
\end{align}
For simplicity we restrict ourselves to only considering decay and neglect dephasing in our calculations unless explicitly mentioned. As we will argue and numerically verify below, the exact nature of the decoherence only plays a minor role for our proposal. The photon loss out of the resonator is described by
\begin{align}
L_{\kappa}=\sqrt{\kappa} a,
\end{align}
where $\kappa$ is the photon loss rate.

Due to our choice of the couplings similar to a $\Lambda$ configuration, most of the dynamics will happen in the two lower levels. To describe them we choose a two-atom basis with triplet states $\ket{00}=\ket{0}_1\ket{0}_2$, $\ket{11}$, $\ket{\rm T}=\frac{1}{\sqrt{2}}\left(\ket{01}+\ket{10}\right)$, and the singlet state $\ket{\rm S}=\frac{1}{\sqrt{2}}\left(\ket{01}-\ket{10}\right)$ as the desired entangled steady state. For the detailed discussion of the engineered decay processes, we also introduce the excited atomic states $\ket{\rm T_0}=\frac{1}{\sqrt{2}}\left(\ket{02} + \ket{20}\right)$, $\ket{\rm S_0}=\frac{1}{\sqrt{2}}\left(\ket{02} - \ket{20}\right)$, $\ket{\rm T_1}=\frac{1}{\sqrt{2}}\left(\ket{12} + \ket{21}\right)$ and $\ket{\rm S_1}=\frac{1}{\sqrt{2}}\left(\ket{12} - \ket{21}\right)$.
The presence of resonator excitations is indicated by a second ket vector, e.g. $\ket{00}\ket{1}$. For simplicity we omit this ket vector when the resonator is in the vacuum state. We use this notation to explain the mechanisms of our scheme in Sec. \ref{SecMechanism} below.

\section{Mechanisms for dissipative preparation of the maximally entangled singlet state}
\label{SecMechanism}

In this section we will show how to engineer effective decay processes which prepare a steady state close to the maximally entangled singlet state $\ket{\rm S}$. For now, we will focus our discussion on the physical mechanisms behind the effective decay processes, while Sec. \ref{SecTwo} and \ref{SecEngineered} will deal with the derivation of quantitative expressions for the effective operators and rates.

The mechanism of our scheme is illustrated in Fig. \ref{FigMechanism} a). The working principle is as follows: Since the singlet state $\ket{\rm S}$ is a dark state of the resonator interaction, it can only gain or loose population by effective decay mechanims mediated by the weak coherent drives or through the slow decay by the weak qubit decoherence. A strong asymmetry between the rapid decay into $\ket{\rm S}$ and the slow loss processes out of it results in the dissipative preparation of $\ket{\rm S}$ with high fidelity. In the following we will discuss the physical mechanism for the preparation of $\ket{\rm S}$.

In the previous section we have introduced a coherent driving $H_{\rm d}$. The purpose of it is to drive a two-photon transition $\ket{0} \leftrightarrow \ket{2}$. For now, we will assume that we have a coherent drive of $\ket{0} \leftrightarrow \ket{2}$ with a coupling constant of $\Omega_{\rm eff}$ and defer the derivation to later. Due to the opposite phase of $\Omega_2$ on the two transmons, this drive then couples $\ket{00}$ to an excited state $\ket{\rm S_0}$ with a detuning of $\delta_2$, as can be seen from Fig. \ref{FigMechanism} a). $\ket{\rm S_0}$ is in turn coupled to $\ket{\rm S}\ket{1}$ by the resonator coupling $H_{\rm cav}$. From here, $\ket{\rm S}\ket{1}$ decays into $\ket{\rm S}$ via resonator decay at a rate of $\kappa$. These processes combine to an effective resonator decay process from $\ket{00}$ into $\ket{\rm S}$ with a rate of $\kappa_+$.

In order to engineer this process to be as strong as possible we have to fulfill two requirements:
First, we need to make sure that the coupling of the transmon-excited state $\ket{\rm S_0}$ to the resonator-excited state $\ket{\rm S}\ket{1}$ is close to resonance, given that only the latter can decay to $\ket{\rm S}$ through resonator photon loss. To this end we set the resonator into or close to resonance with the upper transition of the transmons, $\ket{2} \leftrightarrow \ket{1}\ket{1}$. This is reached by choosing $\omega_{\rm c} = \omega - 2A$ ($\delta_{\rm c} = \delta_2 - \delta_1$), and results in an equal energy of $\ket{\rm S_0}$ and $\ket{\rm S}\ket{1}$, as shown in Fig. \ref{FigMechanism} a). The two states hybridize and form dressed states
\begin{align}
\ket{\rm S_\pm} = \frac{1}{\sqrt{2}} \left(\ket{\rm S_0} \pm \ket{\rm S}\ket{1}\right),
\label{EqDressed}
\end{align}
located at frequencies of $2 \omega - 2 A \pm \sqrt{2} g$ (or $\delta_2 \pm \sqrt{2} g$).

The second requirement is that the two-photon drive from from $\ket{00}$ is resonant with one of the dressed states in Eq. (\ref{EqDressed}). Choosing a detuning of $\delta_2 = \sqrt{2} g$, we tune the drive into resonance with the transition from $\ket{00}$ to $\ket{\rm S_-}$. Population from $\ket{00}$ is then rapidly excited to $\ket{\rm S_-}$, which, through its contribution from $\ket{\rm S}\ket{1}$, decays into $\ket{\rm S}$. For a strong resonant drive, the resulting effective decay process is only limited by the line width $\frac{\kappa}{2}$ of $\ket{\rm S_-}$, the state which mediates it. Thus, the dissipative preparation mechanism of the singlet and its rate $\kappa_+$ can be engineered to be rather large.

Loss from the singlet can occur through the couplings of $\ket{\rm S}$ to any excited state other than $\ket{\rm S_0}$ by the available microwave fields, e.g. to $\ket{\rm T_1}$ by $\Omega_{\rm eff}$. As indicated in Fig. \ref{FigMechanism} c), these excited states are coupled to a number of other, in particular resonator-excited states. For instance $\ket{\rm T_1}$ couples to $\ket{11}\ket{1}$, $\ket{\rm T_0}\ket{1}$, $\ket{\rm T}\ket{2}$, and $\ket{00}\ket{3}$. Consequently, this establishes a loss channel from $\ket{\rm S}$ through effective resonator decay, e.g. into $\ket{11}$, which causes losses at a rate $\kappa_-$ from the desired steady state $\ket{\rm S}$.
Fortunately, the photon-number dependent coupling strength between transmons and resonator provides us with a non-equidistant spectrum which consequently makes it possible to have the two-photon drive resonant with the transition from $\ket{00}$ to $\ket{\rm S_-}$ while keeping it off-resonant with the transitions from $\ket{\rm S}$ to other hybridized excited states. In this way, loss processes from the singlet are suppressed by their detunings.

In order to reach $\ket{\rm S}$ independently from the initial state and to maintain it as the steady state, an additional mechanism is required to transfer population from lower states other than $\ket{00}$, i.e. from $\ket{\rm T}$ and $\ket{11}$, to $\ket{\rm S}$.
So far, we have assumed that the resonator is resonant with the upper transition. This means that due to the anharmonicity, the resonator is off-resonant with the lower transition. For reasonable anharmonicities the off-resonant coupling is, however, still sufficient to allow a reshuffling of population from the bright states $\ket{11}$ and $\ket{\rm T}$ to $\ket{00}$, while $\ket{\rm S}$ as the dark state of the resonator coupling remains unaffected. As is shown in Fig. \ref{FigMechanism} b), this reshuffling process involves the resonator coupling of the lower transition ($\sqrt{2} g$), e.g. $\ket{\rm T} \leftrightarrow \ket{00}\ket{1}$, and decay of a resonator excitation at a rate of $\kappa$. It can be seen as an effective decay process with a decay rate $\kappa_{\rm eff} = 2 \kappa g^2 / [2 g^2 + (\delta_{\rm c}-\delta_1)^2/2 + \kappa^2/4]$. This expression contains both limiting cases, where one can either eliminate the resonator-excited states, or where the states can be seen as dressed states with resonator-excited states, for instance the triplet states
\begin{align}
\ket{\rm T_\pm} = \frac{1}{\sqrt{2}} \left(\ket{\rm T} \pm \ket{00}\ket{1}\right),
\end{align}
which decay towards $\ket{00}$ at rates $\propto \kappa$.
Ideally, the reshuffling mechanism rapidly transfers the population of the triplet states to $\ket{00}$, from where they decay into $\ket{\rm S}$ by the dissipative preparation mechanism discussed above. The fastest reshuffling is reached by tuning the resonator into resonance with the lower transition, i.e. $\delta_{\rm c} = \delta_1$. This choice is, however, different from the above choice of $\delta_{\rm c} = \delta_2 - \delta_1$ which optimizes the dissipative state preparation process.
With this choice of the resonator frequency we get $\kappa_{\rm eff} = 2 \kappa g^2 / [2 g^2 + 2 A^2 + \kappa^2/4]$, from which we see that the reshuffling works best for small anharmonicity $A$. For larger $A$ the process becomes less effective.
Having both processes, state preparation and reshuffling, simultaneously active might therefore seem problematic for large anharmonicities. However, as we shall see below, the scheme can still be effective for large $A$ if we allow for longer time for the reshuffling.
Furthermore, as is also addressed below, the two requirements for $\delta_{\rm c}$ above are far less critical than the resonant set-up of the two-photon drive. Consequently, both processes, the dissipative state preparation and the reshuffling, can be effective at the same time over a wide parameter range, as we will numerically demonstrate in Sec. \ref{SecParameter}.

In addition to effective resonator decay, qubit decoherence present in the system can cause loss from the singlet independent of the drives. Most notably, it can cause a loss from $\ket{\rm S}$ into $\ket{00}$, as shown in Fig. \ref{FigMechanism} c).
The presented mechanisms are summarized in Fig. \ref{FigMechanism} d): On the left hand side we see the reshuffling mechanisms enabled by the resonator coupling to the lower transition, represented by $\kappa_{\rm eff}$, and on the right hand side the state preparation ($\kappa_+$) and loss ($\kappa_-$) mechanisms affecting the singlet state, as well as the decay from $\ket{\rm S}$ by qubit decoherence at a rate of $\gamma$.

To sum up this section, we have identified suitable mechanisms for the dissipative preparation of the singlet state and discussed the physical effects behind them. In the following two sections we will analytically derive the couplings and the rates for the effective coherent and dissipative processes in our scheme. Based on these, we derive benchmarks for the performance of the scheme in Sec. \ref{SecParameter}.

\subsection{Effective coherent driving of the dipole-forbidden transition $\ket{0} \leftrightarrow \ket{2}$ by a two-photon process}
\label{SecTwo}
The implementation of the dissipative state preparation scheme discussed above requires a coherent coupling of the transition $\ket{0} \leftrightarrow \ket{2}$. Since this transition is dipole-forbidden, such a coupling cannot be accomplished in a single step. One way to overcome this is to use a two-photon process, achieved by the combination of two individual fields. In $\hat{H}_{\rm d}$ we have chosen two such fields, $\Omega_1$ and $\Omega_2$. As we will derive in the following, these provide complementary virtual single-photon excitations which form the desired coupling. 

In the following, we will apply the effective operator formalism presented in Ref. [\onlinecite{EO}] to obtain a simple effective Hamiltonian for a single transmon with a two-photon drive. Here, we separate the Hamiltonian into a perturbative part $V(t) = H_{\rm d}$, which contains the fields, and a perturbed part $H_0 = H^{'}_{\rm free} - \delta_c a^\dagger a$. (Note that the derivation below is for a single transmon only. With this in mind, the reuse of Hamiltonian definitions should not cause any confusion.) 
While in Ref. [\onlinecite{EO}] only effective processes with an initial excitation are considered, here we also allow for an initial deexcitation. We therefore set up the effective Hamiltonian (cf. Ref. [\onlinecite{EO}]) as $H_{\rm eff} = H_{\rm eff}^{(+)} + H_{\rm eff}^{(-)}$ with
\begin{align}
\label{effHall}
H_{\rm eff}^{(\pm)} = &- \frac{1}{2} V(t) \sum_{f=1}^2 \sum_{k=0}^2 \left(H_0^{(k, f,\pm)}\right)^{-1} V_\pm^{(k,f)}(t) + H.c.,
\end{align}
Here, we specify the initial state $k$ and the field $f$ of the perturbation $V_\pm^{(k,f)}$ and the unperturbed Hamiltonian $H_0^{(k, f,\pm)}$. The latter is defined as $H^{'}_{\rm free} \pm \Delta_f - \omega_k$ and contains $\omega_k$ as the frequency of level $k \in \{0,1,2\}$ and $\Delta_f$ as the detuning of field $f \in \{1,2\}$. We use a projector $P_k = \ket{k} \bra{k}$ on the levels $k$ to identify coherent drive terms $V_\pm^{(k,f)} = V^{(f)} P_k$ starting from an initial state $k$. The superscript $f \in \{1,2\}$ is used to split $V(t)$ into $V_\pm^{(k,1)}$ for those terms which depend on $\Omega_1$ and $V_\pm^{(k,2)}$ for the ones with $\Omega_2$; a sign $(\pm)$ denotes whether the initial process is an excitation $(+)$, i.e. a term containing a factor $e^{-i \omega_f t}$, or a de-excitation $(-)$, with a factor $e^{+ i \omega_f t}$.

Using this formalism we find a considerable number of terms, time-independent and -dependent ones, some closer to resonance and others stronger detuned. Neglecting the time-varying terms rotating at twice a detuning $\Delta_{1/2}$ we obtain the effective two-photon Hamiltonian
\begin{align}
H_{\rm eff} \approx &\sum_{j=1,2} \sum_{f=1,2} \frac{\Omega_1^2}{4 (\delta_1 + \Delta_f)} \left(\ket{1}_j \bra{1} - \ket{0}_j \bra{0} \right) \\ 
&- \frac{(-1)^j \Omega_1 \Omega_2}{4 \sqrt{2} (\delta_1 + \Delta_f)} \left(\ket{2}_j\bra{0} + \ket{0}_j\bra{2}\right) \nonumber \\
&+ \frac{\Omega_f^2}{2 (\delta_1 - \delta_2 - \Delta_f)} \left(\ket{1}_j \bra{1} - \ket{2}_j \bra{2} \right) \nonumber \\
&- \frac{(-1)^j \Omega_1 \Omega_2}{4 \sqrt{2} (\delta_1 - \delta_2 - \Delta_f)} \left(\ket{2}_j\bra{0} + \ket{0}_j\bra{2}\right) \nonumber
\end{align}
Setting the detunings of the fields to $\Delta_{1/2} = \mp(\delta_1 + \epsilon)$ we have that $\Delta_1 + \Delta_2 = 0$ and keep a certain virtual character of the single fields by a detuning of $\pm \epsilon$, as shown in Fig. \ref{FigSetup} b). In this configuration, there exists an effective two-photon drive where the first field (with $\Omega_1$) drives the lower transition $\ket{0} \leftrightarrow \ket{1}$ and the second field (with $\Omega_2$) drives the upper transition. Expressing the resulting effective Hamiltonian in terms of the anharmonicity (using $\delta_1 = \frac{\delta_2}{2} - A$) we obtain
\begin{align}
H_{\rm eff} \approx &\sum_{j=1,2} \left(\frac{\Omega_1^2}{4 \epsilon} - \frac{\Omega_2^2}{4(2 A + \delta_2 + \epsilon)} \right)\left(\ket{0}_j \bra{0} - \ket{1}_j \bra{1} \right) \nonumber \\ 
&+ \left(- \frac{\Omega_2^2}{2 (\delta_2 + \epsilon)} + \frac{\Omega_1^2}{2(2A+\epsilon)} \right) \left(\ket{1}_j \bra{1} - \ket{2}_j \bra{2} \right) \nonumber \\
&+ \frac{\Omega_{\rm eff}}{2} (-1)^j \left(\ket{2}_j\bra{0} + \ket{0}_j\bra{2}\right) 
\label{EqHeffDrive1}
\end{align}
with an effective two-photon Rabi frequency
\begin{align}
\Omega_{\rm eff} = &\frac{\Omega_1 \Omega_2}{2 \sqrt{2}} \left(\frac{1}{\epsilon} + \frac{1}{\delta_2 + \epsilon} - \frac{1}{2 A + \epsilon} - \frac{1}{2 A + \delta_2 + \epsilon}\right) \nonumber \\
= &\frac{\Omega_1 \Omega_2}{2 \sqrt{2}} \frac{2 A \delta_2 [2 (A - \epsilon) + \delta_2]}{\epsilon (\delta_2 + \epsilon)(2 A + \epsilon)(2 A + \delta_2 + \epsilon)}.
\label{EqEffTwo}
\end{align}
From here we see that for the case of zero anharmonicity $A = 0$, i.e. for harmonic transmons, no effective two photon drive is possible. For $A \neq 0$, however, there exists a possibility of driving the transition $\ket{0} \leftrightarrow \ket{2}$. Note that the remaining diagonal terms in Eq. (\ref{EqHeffDrive1}) represent shifts which can be compensated by suitable (minor) detunings of the fields. Their effect on Eq. (\ref{EqHeffDrive1}) can be considered very small so that $H_{\rm eff}$ is approximately given by a single coherent coupling of the transition $\ket{0} \leftrightarrow \ket{2}$,
\begin{align}
H_{\rm d, eff} = &\sum_{j=1,2} \frac{\Omega_{\rm eff}}{2} (-1)^j \ket{2}_j\bra{0} + H.c.
\label{EqEffDrive}
\end{align}
We have thus obtained the coupling constant $\Omega_{\rm eff}$ of the effective two-photon coupling we introduced in Sec. \ref{SecSetup}. With this result we can turn to the derivation of the effective Lindblad operators for the engineered decay mechanisms used for the preparation of the singlet state.

\subsection{Engineered decay processes and their effective Lindblad operators}

\label{SecEngineered}
To model the effective, dissipative evolution we use the same effective formalism as in the previous section to derive the effective Lindblad operators \cite{EO}
\begin{align}
\label{effLall}
L_{\rm eff}^m &= L_m \sum_{k} \sum_{f} \left(H_{\rm NH}^{(k, f)}\right)^{-1} V^{(k, f)}(t),
\end{align}
with the perturbative coherent excitation $V^{(k, f)}(t)$ from an initial state $k$ by a field $f$, and a non-Hermitian Hamiltonian
\begin{align}
H_{\rm NH}^{(k, f)} = H_0^{(k, f)} - \frac{i}{2} \sum_n L_n^\dagger L_n
\label{EqHNH}
\end{align}
with the perturbed Hamiltonian $H_0^{(k, f)}$ defined previously. We focus on the effective resonator decay process activated by the two-photon drive $H_{\rm eff}$ and followed by decay of a resonator excitation $L_\kappa$. With $H_0 = H^{'}_{\rm free} + H^{'}_{\rm cav}$, $V(t) = H_{\rm eff}$ ($\Omega_{\rm} \ll \delta_2$), and $L_m = L_\kappa$ we arrive at an effective Lindblad operator
\begin{align}
L^{\kappa}_{\rm eff} \approx &\sqrt{\kappa_+} \ket{{\rm S}}\bra{00} + \sum_j \sqrt{\kappa^-_j} \ket{\phi_j}\bra{\rm S},
\label{EqEffCavity1}
\end{align}
with effective decay rates of $\kappa_+$ and $\kappa^-_j$. This operator represents the dissipative mechanism we engineer to rapidly prepare the singlet state $\ket{\rm S}$ from $\ket{00}$. In addition, it includes the loss processes at rates of $\kappa^{-}_j$ from $\ket{\rm S}$ into other states $\ket{\phi_j} \in \{\ket{11}, \ket{\rm T_0}, \ket{\rm T,1}, \ket{00,2}\}$. Note that here we have ignored some less important terms as their effect on the population of the singlet is small.

We calculate $\kappa_+$ of Eq. (\ref{EqEffCavity1}), using the driving from $\ket{00}$ to $\ket{\rm S_0}$ as given by Eq. (\ref{EqEffDrive}), with a matrix element of $\frac{\Omega_{\rm eff}}{\sqrt{2}}$. The dynamics of the excited state $\ket{\rm S_0}$ is described by the non-Hermitian Hamiltonian in Eq. (\ref{EqHNH}) which couples $\ket{\rm S_0}$ to $\ket{\rm S}\ket{1}$ through the resonator interaction $H^{'}_{\rm cav}$, forming a coupled subspace. For the non-Hermitian Hamiltonian $H_{\rm NH}^{(\rm \ket{00}, \Omega_{\rm eff})}$ of this subspace which contains $\ket{\rm S_0}$ and is reached by excitation from $\ket{\rm S}$ with the two-photon drive $H_{\rm eff}$, we define $H_{\rm S_0} \equiv H_{\rm NH}^{(\rm \ket{00}, \Omega_{\rm eff})}$ with
\begin{align}
H_{\rm S_0} = ~ &\tilde{\delta}_2 \ket{\rm S_0}\bra{\rm S_0} + (\tilde{\delta}_1 + \tilde{\delta}_{\rm c}) \ket{\rm S}\ket{1}\bra{1}\bra{\rm S} + \nonumber \\ &+ \sqrt{2} g \left(\ket{\rm S}\ket{1}\bra{\rm S_0} + H.c.\right).
\end{align}
In order to keep the notation compact, we have written the Hamiltonian in terms of the complex detunings $\tilde{\delta}_j = \delta_j - \frac{i j \gamma}{2}$ and $\tilde{\delta}_{\rm c} = \delta_{\rm c} - \frac{i \kappa}{2}$ combining the energy with the imaginary line width of the levels. For the inverted operator we find
\begin{align}
H_{\rm S_0}^{-1} = ~ &\tilde{\delta}^{-1}_{2, \rm eff} \ket{\rm S_0}\bra{\rm S_0} + \tilde{\delta}^{-1}_{\rm 1c, eff} \ket{\rm S}\ket{1}\bra{1}\bra{\rm S} + \nonumber \\ &+ \tilde{g}^{-1}_{\rm eff} \left(\ket{\rm S}\ket{1}\bra{\rm S_0} + H.c.\right).
\end{align}
Here, we have introduced effective detunings of $\delta_{\rm 2, eff} = \tilde{\delta}_2 - \frac{2g^2}{\tilde{\delta}_2}$ and $\delta_{\rm 1c, eff} = (\tilde{\delta}_1 + \tilde{\delta_{\rm c}}) - \frac{2g^2}{\tilde{\delta}_1 + \tilde{\delta_{\rm c}}}$, and an effective coupling constant of $\tilde{g}_{\rm eff} = \sqrt{2} g - \frac{\tilde{\delta}_2 (\tilde{\delta}_1 + \tilde{\delta}_{\rm c})}{\sqrt{2} g}$. Since the rate for resonator decay from $\ket{\rm S}\ket{1}$ into $\ket{\rm S}$ is given by $\kappa$, we generally find an effective decay of $\kappa_+ = \frac{\kappa \Omega_{\rm eff}^2}{2 |\tilde{g}_{\rm eff}|^2}$ from $\ket{00}$ to $\ket{\rm S}$, concluding that the effective coupling rate $\tilde{g}_{\rm eff}$ governs the strength of the engineered decay process.

The decay rate $\kappa_+$ is maximized by a parameter choice of $\delta_2 = \sqrt{2} g$ and $\delta_{\rm c} = \delta_2 - \delta_1$, which corresponds to the two-photon drive from $\ket{00}$ being in resonance with $\ket{\rm S_0}$ and the resonator being resonant with the upper transition. We then obtain $\tilde{g}_{\rm eff} \approx \frac{i \kappa}{2}$, and thus $\kappa_+ \approx \frac{\Omega_{\rm eff}^2}{\kappa}$. In Sec. \ref{SecParameter} we will make use of this result to derive the error and the speed of the protocol.

We now turn to the effective loss processes $\kappa^-_j$ as they appear in Eq. (\ref{EqEffCavity1}).
Given that $\ket{\rm S}$ is a dark state of the resonator coupling, these rates can be calculated using the same procedure we applied for the derivation of $\kappa_+$ above: As $\ket{\rm S}$ is coupled to $\ket{\rm T_1}$ by the two-photon drive we need to consider the non-Hermitian Hamiltonian $H_{\rm T_1} \equiv H_{\rm NH}^{(\rm \ket{\rm S}, \Omega_{\rm eff})}$ which describes the subspace consisting of $\ket{\rm T_1}$ and the states coupled to it by $H^{'}_{\rm cav}$.
For low anharmonicities $A \lesssim \delta_2$, $H_{\rm NH, T_1}$ needs to reflect the full complexity of the coupled subspace containing $\ket{\rm T_1}$, $\ket{11}\ket{1}$, $\ket{\rm T_0}\ket{1}$, $\ket{\rm T}\ket{2}$ and $\ket{00}\ket{3}$.
For anharmonicities of $A \gtrsim \delta_2$, however, the subspace of $\ket{\rm T_1}$ and $\ket{11}\ket{1}$ begins to decouple from the other states so that the dynamics of the excited states can be approximated using only $\ket{\rm T_1}$ and $\ket{11}\ket{1}$.
The Lindblad operator of Eq. (\ref{EqEffCavity1}) for the effective resonator decay then reduces to
\begin{align}
L^{\kappa}_{\rm eff} \approx &\sqrt{\kappa_+} \ket{{\rm S}}\bra{00} + \sqrt{\kappa_-} \ket{11}\bra{\rm S},
\label{EqEffCavity2}
\end{align}
containing a single loss rate $\kappa_- = \kappa^-_{\ket{11}}$ from $\ket{\rm S}$ into $\ket{11}$.

To derive $\kappa_-$, we then approximate $H_{\rm NH, T_1}$ by the non-Hermitian Hamiltonian of the excited subspace consisting of $\ket{\rm T_1}$ and $\ket{11}\ket{1}$,
\begin{align}
H_{\rm T_1} \approx ~ &\tilde{\delta}_2 \ket{\rm T_1}\bra{\rm T_1} + (\tilde{\delta}_1 + \tilde{\delta}_{\rm c}) \ket{11}\ket{1}\bra{1}\bra{11} + \nonumber \\ &+ 2 g \left(\ket{\rm 11}\ket{1}\bra{\rm T_1} + H.c.\right),
\end{align}
using the complex detunings defined above.
The inverted operator is then given by
\begin{align}
H^{-1}_{\rm T_1} \approx ~ &\tilde{\delta}^{-1}_{2, \rm eff} \ket{\rm T_1}\bra{\rm T_1} + \tilde{\delta}^{-1}_{\rm 1c, eff} \ket{11}\ket{1}\bra{1}\bra{\rm T} + \nonumber \\ &+ \tilde{g}^{-1}_{2, \rm eff} \left(\ket{11}\ket{1}\bra{\rm T_1} + H.c.\right).
\end{align}
Here, we have found effective detunings $\delta_{\rm 2, eff, T_1} = \tilde{\delta}_2 - \frac{4 g^2}{\tilde{\delta}_2}$ and $\delta_{\rm 1c, eff, T_1} = (\tilde{\delta}_1 + \tilde{\delta_{\rm c}}) - \frac{4 g^2}{\tilde{\delta}_1 + \tilde{\delta_{\rm c}}}$, and an effective coupling constant of $\tilde{g}_{\rm eff, T_1} = 2 g - \frac{\tilde{\delta}_2 (\tilde{\delta}_1 + \tilde{\delta}_{\rm c})}{2 g}$, which are different from the ones in the previous case of $\ket{\rm S_0}$. With a decay rate $\kappa$ from $\ket{\rm 11}\ket{1}$ into $\ket{\rm 11}$, we obtain an effective decay rate of $\kappa_- \approx \frac{\kappa \Omega_{\rm eff}^2}{|\tilde{g}_{\rm eff, T_1}|^2}$ for the losses from $\ket{\rm S}$.
For the above choice of $\delta_2$ and $\delta_{\rm c}$, the effective coupling constant becomes $\tilde{g}_{\rm eff, T_1} \approx g$ which results in $\kappa_- \approx \frac{\kappa \Omega_{\rm eff}^2}{4 g^2}$. From here we conclude that for $\kappa^2 \ll g^2$ the effective loss rate $\kappa_-$ from the singlet is engineered to be much smaller than its preparation rate $\kappa_+ \approx \frac{\Omega_{\rm eff}^2}{2 \kappa}$. These results confirm the explanations in Sec. \ref{SecMechanism}.

Note that, on the one hand, the above treatment of the coupled excited subspace where we restrict the excited state subspace to $\ket{\rm T_1}$ and $\ket{11}\ket{1}$ is quite simplistic, given that it reduces the number of resonances from five to only two.
In particular, one needs to ensure that one does not hit an accidental resonance with one of the dressed states of the system.
On the other hand, the parameter space consisting of $\delta_{\rm c}$, $\delta_2$ and $\epsilon$ is sufficiently big to avoid an excitation of the remaining undesired resonances as there are sufficiently many suitable points in different regions of parameter space for which all of these resonances are off-resonant with the two-photon drive.
\begin{figure}[t]
\centering
\includegraphics[width=7.0cm]{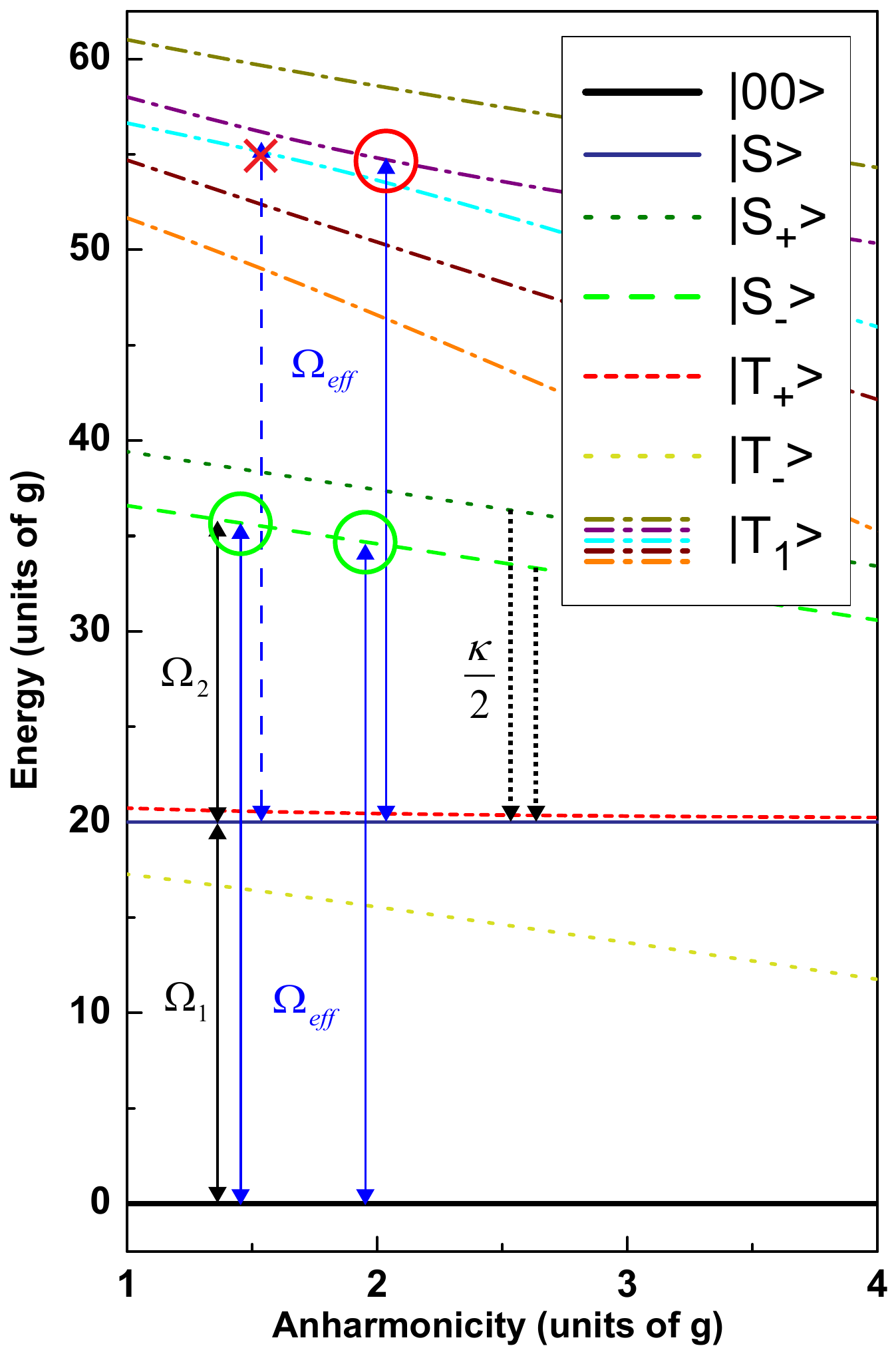}
\caption{(Color). Dressed state energy vs. anharmonicity. An effective two-photon drive $\Omega_{\rm eff}$ from $\ket{00}$ (solid black line) to $\ket{\rm S_-}$ (green dashed) is implemented as two consecutive single-photon excitations by two microwave fields, $\Omega_1$ and $\Omega_2$. The individual drives are mediated by $\ket {\rm T_+}$ (short-dashed red), which is a dressed state of $\ket{\rm T}$, and made virtual through a detuning of $\epsilon$ (not shown).
The two virtual excitations combine to an effective drive $\Omega_{\rm eff}$ resonant with the transition $\ket{00} \leftrightarrow \ket{\rm S_-}$; $\ket{\rm S_-}$ then decays into $\ket{\rm S}$ (indicated). The same field couples to the transition from $\ket{\rm S_-}$ to the dressed states of $\ket{\rm T_1}$ (dashed-dotted).
By an appropriate choice of the oscillator detuning $\delta_{\rm c}$ (here plotted for $\delta_{\rm c} = \delta_2 - \delta_1$ with $\omega = 20 g$), this coupling to $\ket{\rm T_1}$ is made off-resonant (left set of arrows). In case that $\ket{\rm T_1}$ is hit by the drive (right set of arrows), $\delta_{\rm c}$ needs to be chosen differently to make the coupling off-resonant.
}
\label{FigDressed}
\end{figure}
In Fig. \ref{FigDressed}, we draw the dressed states of the coupled resonator-transmon system. Here, the single-photon fields are tuned to resonantly excite the transition $\ket{00} \leftrightarrow \ket{\rm S_-}$ by a two-photon transition, mediated by the triplet state $\ket{\rm T}$. The same two-photon drive also couples $\ket{\rm S}$ to a number of dressed states with contributions from $\ket{\rm T_1}$. These transitions, however, generally have different frequencies than the desired one from $\ket{00}$ to $\ket{\rm S_-}$ so that excitation of $\ket{\rm S}$ by the drive $\Omega_{\rm eff}$ is off-resonant and suppressed by its detuning from the dressed states. This can be seen from Fig. \ref{FigDressed}, where we draw the dressed states together with the two-photon drive for the choice of $\delta_{\rm c} = \delta_2 - \delta_1$. 
In the figure, we show an example near $A = \frac{3 g}{2}$ where the driving is off-resonant with the excited states which contain contributions from $\ket{\rm T_1}$. We also draw an example at $A \approx 2 g$ where this is not the case and where a resonance is hit accidentally. Here, it is necessary to choose a different detuning $\delta_{\rm c}$.
Below, we will verify by numerical simulation for a broad parameter range that it is always possible to avoid such resonances.

In addition to losses caused by the two-photon drive, also the individual fields $\Omega_1$ and $\Omega_2$ couple $\ket{\rm S}$ to other states. The coupling of the even-phase single-photon drive $\Omega_1$ from $\ket{\rm S}$ to $\ket{\rm S_0}$ does not cause any significant loss from $\ket{\rm S}$, since population in $\ket{\rm S_0}$ is recycled via $\ket{\rm S}\ket{1}$ back into $\ket{\rm S}$. The odd-phase single-photon drive $\Omega_2$, on the other hand, couples $\ket{\rm S}$ to $\ket{00}$ and to a superposition state $\frac{1}{\sqrt{2}}(\ket{11}-\ket{\rm T_0})$. Both these states are dark states of the resonator coupling. Thus, no exchange excitation to a resonator-excited state can shift them into resonance with the off-resonant drive $\Omega_2$ from $\ket{\rm S}$ and no effective resonator decay process from $\ket{\rm S}$ is established involving them. Accumulation in these states does not occur, either, given that $\frac{1}{\sqrt{2}}(\ket{11}-\ket{\rm T_0})$ decays through qubit decoherence and $\ket{00}$ decays into $\ket{\rm S}$ as discussed earlier.
As a consequence, neither of the two drives causes significant loss from the singlet.

Another source of errors emerges for small anharmonicities $A \lesssim \delta_2$ from the coherent coupling of $\ket{\rm S}$ to other states like $\ket{00}$ by the single-photon drives $\Omega_1$ and $\Omega_2$. However, for $A \gtrsim \delta_2$, these couplings are sufficiently detuned to be ignored. Also, beside effective resonator decay processes, qubit decoherence occurs according to Eqs. (\ref{EqSpont1})-(\ref{EqSpont2}). Provided that the decay rate $\gamma$ is much weaker than all other physical couplings present in the system, i.e. $\gamma \ll \kappa, g$, effective processes combining qubit decoherence with coherent excitation can be safely neglected.

We conclude that the sources of error originating from effective resonator decay which can cause losses from the singlet state are suppressed for the right parameter choice. These processes are, together with the engineered dissipative state preparation process, contained in the effective resonator decay operator in Eq. (\ref{EqEffCavity2}).

\section{Parameter and performance analysis, imperfections and realization aspects}
\label{SecParameter}
In the previous section we have identified the effective coherent and dissipative processes which are relevant for our dissipative state preparation scheme and investigated the corresponding Lindblad operators and rates. In this section, we will use these results to derive approximate expressions for the error and speed of the presented protocol as the main benchmarks for our scheme. Later, we will assess the temporal evolution of the system numerically.

\subsection{Error and speed of the protocol}
In the previous section we have derived the effective resonator decay operator $L^\kappa_{\rm eff}$, given in Eq. (\ref{EqEffCavity2}), which describes both the preparation of the singlet state $\ket{\rm S}$ and the inherent losses of our scheme. The derivation of Eq. (\ref{EqEffCavity2}) was carried out in the limit of weak driving.
As we will find numerically below, the dissipative preparation of the singlet at a rate of $\kappa_+ \approx \frac{\Omega_{\rm eff}^2}{2 \kappa}$ works well for a driving strength up to $\Omega_{\rm eff} \approx \frac{\kappa}{8}$, which yields a preparation rate $\kappa_+ \approx \frac{\kappa}{128}$ for the singlet state $\ket{\rm S}$ and a loss rate $\kappa_- \approx \frac{\kappa^3}{256 g^2}$ from it.
In addition, $\ket{\rm S}$ decays at a rate of $\gamma$, as described by the operators in Eq. (\ref{EqSpont1})-(\ref{EqSpont2}).

\begin{figure}[t]
\centering
\includegraphics[width=\columnwidth]{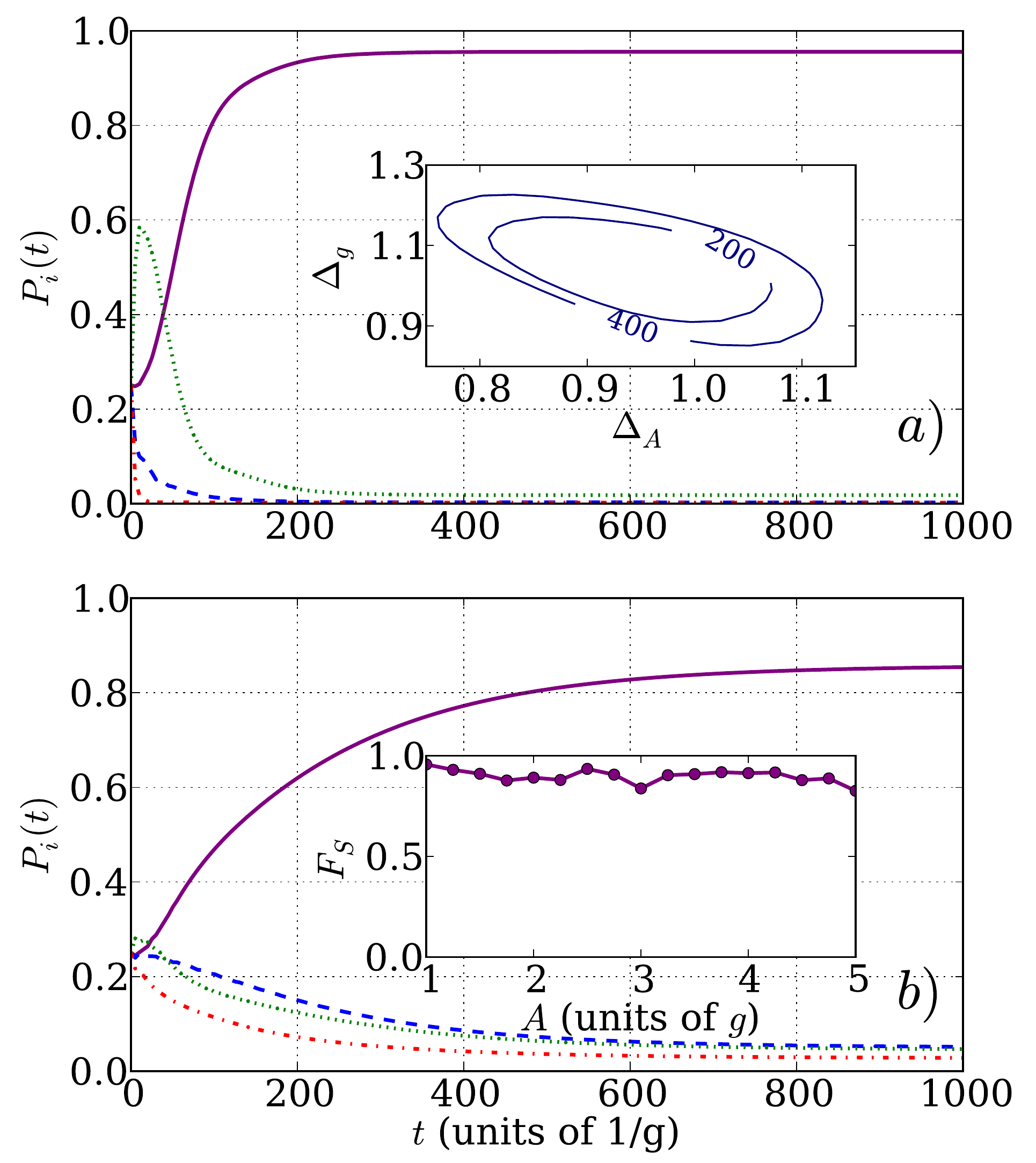}
\caption{(Color) Evolution of the system towards an entangled steady state. Initially prepared in an equal mixture of the lower states ($\ket{00}$ -- green, dotted line, $\ket{11}$ -- red, dashed-dotted line, $\ket{\rm T}$ -- blue, dashed line, $\ket{\rm S}$ -- purple, solid line) the system evolves towards its steady state which is close to the maximally entangled singlet state of the two transmons. Part a) and b) show the result for an anharmonicity of $A = g$ and $A = 4.75g$ respectively. The remaining parameter values are $\Omega_{1/2} = g/3$, $\kappa = 3g/10$ and $\gamma = g/5400$ for all plots. The values of $\bar{\omega}$, $\Delta_{1/2}$ and $\delta_c$ are obtained through numerical optimization. The inset in a) shows the region in the $\Delta_A - \Delta_g$ plane where the singlet fidelity is high, $F_{\rm S} > 90 \%$, for $A = g$. The number on each contour line indicates the preparation time in units of $1/g$. The inset in b) shows the singlet state fidelity at $t = 1000/g$ as a function of anharmonicity.}
\label{FigEvolution}
\end{figure}

Based on these rates we can approximate the temporal dynamics for weak driving using rate equations of the populations $P_i \equiv \bra{\psi_i} \rho \ket{\psi_i}$.
We assume that the reshuffling mechanism rapidly transfers all population from the triplet states to the state $\ket{00}$, which is correct for small anharmonicity $A$, the evolution of the population of the singlet can then be described by a single rate equation for the population of the singlet $P_{\rm S}$,
\begin{align}
\dot{P}_{\rm S} = \kappa_+ P_{00} - (\kappa_- + \gamma) P_{\rm S},
\label{EqRate}
\end{align}
formulated in terms of the decay rates specified above.
Note that in this limit it is only the total decay rate out of the singlet state which matters, since any population lost from it is rapidly reshuffled to the $\ket{00}$ state regardless of the nature of the loss. Hence additional decoherence mechanisms, e.g. dephasing causing decay from $\ket{\rm S}$ to $\ket{\rm T}$, can easily be incorporated be replacing $\gamma$ by an appropriate total loss rate from the singlet.
By simply comparing the gain and loss of the singlet in the steady state, i.e. $\dot{P}_{\rm S} = 0$, we can estimate the steady-state fidelity $F_{\rm S} = \lim\limits_{t \rightarrow \infty}{P_{\rm S}}$ of the singlet and, consequently, the error of the protocol $(1-F_{\rm S})$. Assuming a near unit fidelity we obtain
\begin{align}
(1-F_{\rm S}) \approx \frac{\gamma + \kappa_-}{\kappa_+} = \frac{128 \gamma}{\kappa} + \frac{\kappa^2}{2 g^2}.
\label{EqError}
\end{align}
From this expression we can readily see that the error of the protocol has a promising scaling with the physical parameters. Specifically, the error depends on the ratios of coupling and noise, $g/\kappa$ and $\kappa/\gamma$ so that it will be small for strong coupling, $g^2 \gg \kappa^2$, and modest qubit decoherence, $\gamma \lll \kappa$.
Under the assumption that we can vary the resonator decay rate $\kappa$ we can minimize the error in Eq. (\ref{EqError}) by choosing $\kappa$. Considering $\frac{\partial}{\partial \kappa} (1-F_{\rm S}) = 0$, we derive the optimal resonator decay rate $\kappa_{\rm opt} = 4\sqrt[3]{2 \gamma g^2}$. Inserting this yields the optimized error of the protocol,
\begin{align}
(1-F_{\rm S})_{\rm opt} \approx 24\left({\frac{2 \gamma}{g}}\right)^{2/3}.
\label{EqErrorOpt}
\end{align}
From here we conclude that for $\gamma \lll g$ the inherent error of the protocol can be limited to very small values. We will later confirm this finding numerically.

In addition, the convergence time, i.e. the decay time of the undesired states, can be approximated using Eq. (\ref{EqRate}), assuming rapid reshuffling of the undesired states to $\ket{00}$. Given that here the preparation of the singlet at a rate $\kappa_+$ is the dominant process, the convergence time $\tau$ for weak driving is given by
\begin{align}
\tau \approx \kappa_+^{-1} \approx \frac{32}{\sqrt[3]{2 \gamma g^2}} ,
\label{EqSpeed}
\end{align}
where we have used $\Omega_{\rm eff} \approx \frac{\kappa_{\rm opt}}{8}$ and $\kappa_{\rm opt}$ from above.

Note that the above expressions for the error and the convergence time are approximate and are derived using results obtained for the assumption of weak driving in Sec. \ref{SecEngineered}. In our numerical simulations below we will optimize a number of parameters including the driving strength to achieve highly entangled states within a preparation time as short as possible. In doing so, we arrive at particular choices of the available parameters which allow us to achieve high fidelities in short time. As these optimal parameters are in a regime where the effective Lindblad operators no longer accurately describe the dynamics \cite{EO}, the findings of Eqs. (\ref{EqError})-(\ref{EqSpeed}) deviate from the simulation results below.

\subsection{Numerical results}
To verify the findings above as well as to investigate the limitations of the approximation we now depart from the analytical treatment in the previous sections and assess the performance of the scheme numerically \cite{QuTIP}. To this end we integrate the master equation in Eq. (\ref{EqMaster}) including the three lowest levels of each transmon, $\ket{0}$, $\ket{1}$ and $\ket{2}$, considered in the analytics, as well as the fourth level of each transmon, $\ket{3}$, and up to three photons in the resonator. While level $\ket{3}$ already has a minor effect, the effect of higher excitations is expected to be negligible. Due to the Stark shifts induced by the driving, we have numerically optimized the sum- and difference frequencies $\bar{\omega}$ and $\Delta_{1/2}$ of the drives, as well as the resonator frequency $\delta_{\rm c}$. In Fig. \ref{FigEvolution} we plot the populations 
\begin{align}
P_i(t) = \mathrm{Tr}\left(\left( |\psi_i\rangle\langle \psi_i| \otimes 1_\mathrm{cav}\right)\rho(t) \right)
\end{align} 
between the time evolved density matrix $\rho(t)$ and the four lower states $\ket{\psi_i} = \ket{00}, \ket{11}, \ket{\rm S}, \ket{\rm T}$ introduced in Sec. \ref{SecSetup}. The results of our simulation are shown in Fig. \ref{FigEvolution} a)-b), where we plot the populations, starting with an initially equal mixture of all four lower states. In Fig. \ref{FigEvolution} a), we consider a rather low anharmonicity $A = g$, which is also what is typically used in experiments \cite{Rigetti, Paik, Sears}.
Here, the population of the states $\ket{11}$ and $\ket{\rm T}$ show a fast drop due to the reshuffling into $\ket{00}$. At the same time, albeit on a slightly longer timescale, the dissipative preparation of the singlet is performed, reaching a fidelity of $90 \%$ within a time of about $\tau \approx 200/g$, and a steady state fidelity of $\sim 96\%$. For a transmon experiment with $g/(2 \pi) = 300 \ {\rm MHz}$ this would allow preparation times of about $\tau \approx 80 \ {\rm ns}$. 
For the results in Fig. \ref{FigEvolution} we have chosen $\gamma/(2 \pi) \approx 60 \ {\rm kHz} \approx g/(2 \pi 5400)$ corresponding to a relaxation time of $T_1 \approx 3 \ {\rm \mu s}$ \cite{Houck} for the above parameter choice. 
This is much shorter than current state-of-the-art 3D transmon qubits where decoherence times of up to $T_2 \sim 95 \ {\rm \mu s}$ and $T_1 \sim 70 \ {\rm \mu s}$ \cite{Chow, Rigetti} have been measured. To accurately simulate this situation we include decay and dephasing rates corresponding to the decoherence times and find that with the numbers for 3D transmons it is possible to reach a steady state fidelity of $\sim 97\%$ for $A = g$. Our analytical results (excluding the negligible effect of pure dephasing) suggest that fidelities of $\gtrsim 99 \%$ can be achieved for $T_1 \gtrsim 150 \ {\rm \mu s}$ (or, in the presence of dephasing, for a corresponding $T_2$ time).
The numbers for the transmon decoherence may, however, be somewhat lower than $70 \ {\rm \mu s}$ in the described circuit QED setup, where two qubits need to be tuned into resonance. In the numerical assessment of our scheme we therefore chose to work with a shorter coherence time of $3 \ {\rm \mu s}$ for the transmon relaxation time, comparable to the coherence time obtained for 2D transmons. In doing so we show the robustness of our scheme against such imperfections as well as the possibility to demonstrate a maximally entangled steady state not only in state-of-the-art 3D, but also in the more commonly used 2D transmon systems.

\subsection{Anharmonicity of the transmon}

As discussed in the previous sections, the coupling of the resonator to the $\ket{0} \leftrightarrow \ket{1}$ transition for each transmon contributes to the scheme by reshuffling the unwanted populations to $\ket{00}$. This coupling, however, gets increasingly detuned for higher anharmonicities $A$. In Fig. \ref{FigEvolution} b) we show the effect of an increasing $A$ on the preparation scheme.
Here, for a rather high anharmonicity of $A = 4.75 g$, the reshuffling of the states $\ket{11}$ and $\ket{\rm T}$ to $\ket{00}$ is slowed down as compared to the result for $A = g$ in Fig. \ref{FigEvolution} a). This can be seen from the drop in the population of $\ket{\rm T}$ and $\ket{11}$ which is much less pronounced in b) than in a). In addition, we observe an increase in the steady state populations of these states.
It is therefore advantageous to work with a rather low anharmonicity, where the coupling to the lower transition is still effective. Such anharmonicities are typical for state-of-the-art experiments \cite{Rigetti, Paik, Sears}.

In the following, we will assess the possibility to operate our scheme for a broader range of anharmonicities, despite the breakdown of the reshuffling. To this end we allow for a rather long preparation time $t = 1000/g$. In the inset in Fig. \ref{FigEvolution} b) we show results achieved using a numerical optimization routine to optimize the fidelity by fine-tuning the frequencies of the microwave fields and the resonator. These degrees of freedom in the parameter choice are used by the optimization routine to avoid undesired resonances by a slight departure from the resonance conditions of the previous sections.
The range of our protocol is then limited by the breakdown of the reshuffling to $A \lesssim 4 g$, as well as to $A \gtrsim g$. For lower $A$ the effective two-photon drive becomes ineffective and couplings to higher levels of the transmons add shifts to the resonances required for the state preparation mechanism. To reach a high fidelity $F_{\rm S} > 90 \%$ of the steady state one should therefore work with anharmonicities between $A \approx g$ and $A \approx 4 g$.

Finally, we briefly comment on the possibility for dissipative state preparation with even more anharmonic systems: In this case we choose to have the resonator in (or close to) resonance with the upper transition. Consequently, the lower transition is largely detuned and its effect negligible. We thereby achieve a situation which is very similar to optical cavity QED with atomic $\Lambda$ schemes -- a system where various schemes for dissipative preparation of entanglement are available \cite{KRS, RKS}. These schemes can then be mapped to the highly anharmonic circuit QED setup. In those schemes the role of the far-detuned resonator coupling on the lower transition is accomplished by an additional microwave field which takes over the reshuffling of the triplet states. In this way, preparation of a steady state close to the maximally entangled singlet state can be achieved for any anharmonicity.
For low anharmonicitiy, however, the coupling of the resonator to the lower transition allows us to avoid this field and thus to simplify the experimental implementation.

\subsection{Experimental imperfections}
From the previous discussion it is clear that our scheme relies on the fact that the two transition frequencies of the transmons are identical. Moreover, we have so far only considered the case when the coupling, $g$, is identical for both transmons. In this section, we depart from these assumptions and consider the effect of experimental imperfections. The transmons are characterized by their spectrum which is set by the effective Josephson energy, $E_J$ and the charging energy $E_C = 2A$ \cite{Koch}. Here, we assume that both $\omega = \sqrt{8E_JE_C} - E_C$ and the anharmonicity differ between the transmons. We also consider the possibility of having different couplings to the resonator.  In Fig. \ref{FigEvolution}a, we  focus our analysis on the charging energy (anharmonicity) and the couplings by considering $A_2= \Delta_A A_1$ and $g_2 = \Delta_g g_1$ where the subscript denotes transmon number. In the inset of Fig. \ref{FigEvolution}a, we plot the region in the $\Delta_A-\Delta_g$ plane where $F > 90\%$ for $A_1 = g$. The different contours correspond to the indicated preparation time and we see that there is roughly a $10 - 20\%$ error tolerance built into the system with respect to these parameters. The reproducibility of $E_C$ and $g$ is set by the precision of the e-beam lithography process and these tolerances are well within the limits of current technology. \\

\begin{figure}[t]
\centering
\includegraphics[width=\columnwidth]{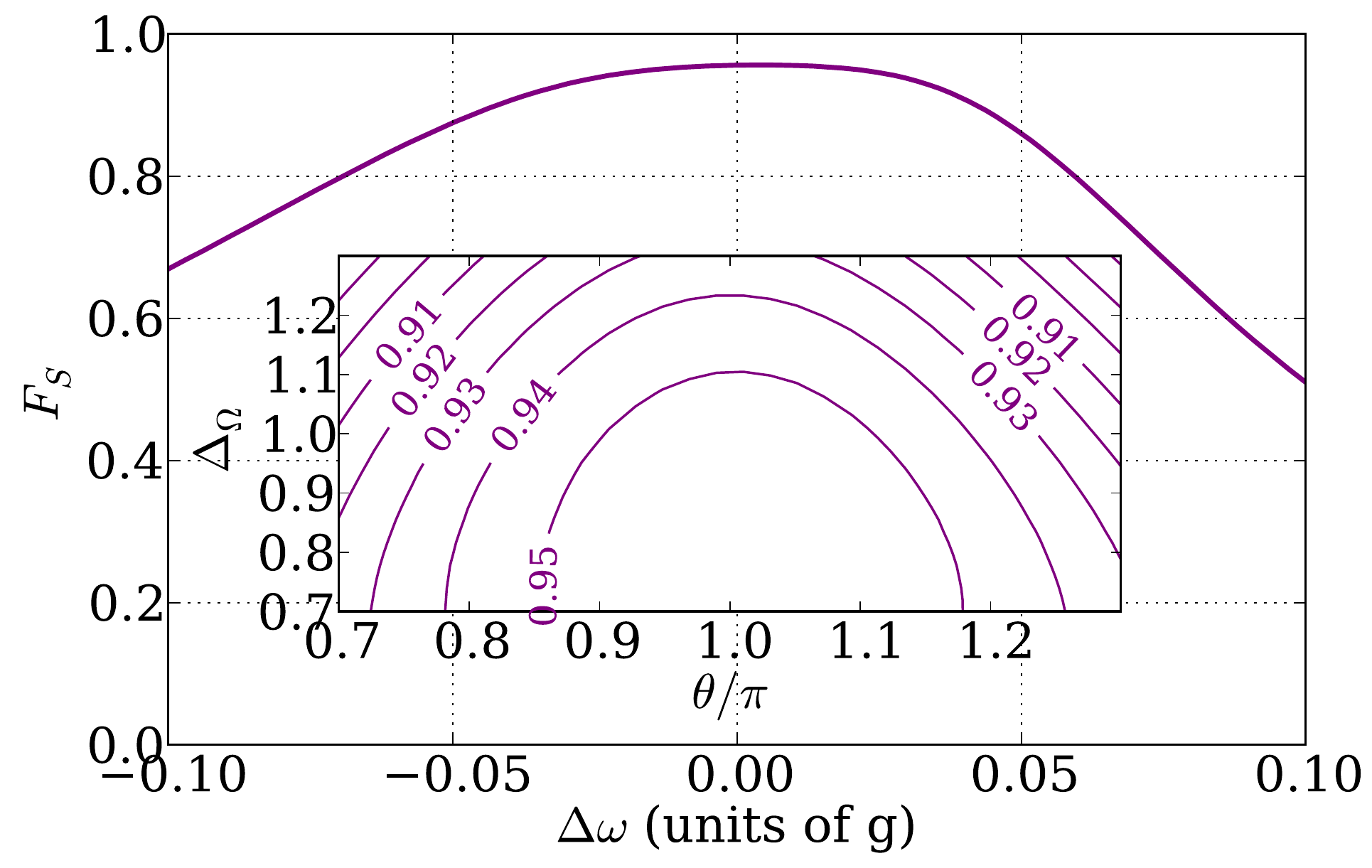}
\caption{(Color) The fidelity as a function of the difference in resonance frequency $\Delta\omega$ between the two transmons. The parameters are as in Fig. \ref{FigEvolution} with $t = 400/g$ and $A = g$. The inset shows the fidelity when varying the amplitude and phase of the microwave signals.}
\label{FigOmegaAndDrive}
\end{figure}

In Fig. \ref{FigOmegaAndDrive}, we consider the effect of different resonance frequencies, $\omega_2 = \omega_1 + \Delta\omega$, where subscripts denote transmon number. The error tolerance with respect to this parameter is substantially smaller than that for differences in anharmonicity and coupling. We believe that this larger sensitivity is due to the fact that for $\omega_1\neq\omega_2$ there is no longer an exact dark state of the transmon-resonator system, and the singlet state begins to suffer from the Purcell enhanced decay, which far exceeds the intrinsic decay rates of the qubits. It is however not necessary to have $\omega$ the same for the two transmons and the tolerance is well within reach of transmon experiments of today.

Apart from differences in circuit parameters, experimental imperfections can also originate from errors in the amplitudes and phases of the continuous microwave tones used to realize the engineered environment. To estimate the robustness of the scheme against such imperfections we consider the drive Hamiltonian 
\begin{align}
H^{'}_{\rm d} = & 
\left(\frac{\Omega_1}{2} e^{i \Delta_1 t} + e^{-i \theta} \frac{\Omega_2}{2} e^{i \Delta_2 t} \right)  \left(\ket{1}_1\bra{0} + \sqrt{2} \ket{2}_1\bra{1}\right) \nonumber \\
 + & 
 \left(\frac{\Omega_1}{2} e^{i \Delta_1 t} + \Delta_\Omega \frac{\Omega_2}{2} e^{i \Delta_2 t} \right)  \left(\ket{1}_2\bra{0} + \sqrt{2} \ket{2}_2\bra{1}\right).
\end{align}
In the inset of Fig. \ref{FigOmegaAndDrive}, we plot the fidelity as a function of $\Delta_\Omega$ and the phase $\theta$. It is clear that there is a substantial robustness in the scheme against imperfections in the microwaves so that no involved tuning scheme is required. We note that the maximum fidelity is not obtained for $\Delta_\Omega = 1$, which indicates that it is in principle possible to optimize all parameters including $\Delta_\Omega$ to achieve even higher values of $F_{\rm S}$.

A different requirement needs to be imposed on the average number of residual thermal photons in the resonator $\bar{n}$. In the absence of residual photons, the target state $\ket{\rm S}$ is a dark state. The preparation of $\ket{\rm S}$ from $\ket{00}$, however, involves a coherent coupling of $\ket{\rm S_0}$ and $\ket{\rm S}\ket{1}$. The singlet is therefore not a dark state in the presence of photons in the resonator which causes a decrease of fidelity for nonzero occupancy numbers, $\bar{n} > 0$. Still, as our numerical simulations show, fidelities of above $90 \%$ are achieved for $\bar{n} \leq 0.02$, a value which is experimentally feasible as demonstrated in Ref. \cite{Sears}. 

\section{Conclusion and outlook}
In this work we have presented a scheme for the preparation of an entangled steady state of two transmons by means of dissipation. We have engineered effective decay mechanisms for the dissipative preparation of the desired maximally entangled singlet state and verified them analytically and numerically. We have demonstrated that high fidelity with the singlet state can be reached within favorable time for realistic experimental parameters, both with 2D and 3D transmons. In addition, our scheme has proven to be robust against experimental imperfections such as non-degeneracy of the transmon levels and couplings.

We consider our proposal for the generation of a small scale entangled state to be a first step towards more advanced protocols in the framework of dissipative state engineering and dissipative quantum computation implemented in superconducting systems. We hope that our scheme will find application in the generation of high-fidelity steady state entanglement in circuit QED setups and that this will stimulate further investigations aiming to harness dissipation for large scale quantum information processing.

\textit{Note added}. Recently, our attention was drawn to the submission of a study [\onlinecite{Leghtas}] with a similar objective. Contrary to our scheme, this proposal works with two two-level systems in the highly dispersive regime. Furthermore, it relies on the frequency difference of the two qubits for breaking the symmetry between the two transmons, whereas our scheme does this by having a different phase on one of the driving fields. The scheme involves six microwave drives as opposed to the four drives in our proposal.

\section*{Acknowledgments}
We thank Jonas Bylander and Per Delsing, as well as Gerhard Kirchmair, Shyam Shankar and Steve Girvin for discussions. The research leading to these results has received funding from the European Research Council under the European Union's Seventh Framework Programme (FP/2007-2013), through the ERC Grant Agreement n. 306576, the Villum Kann Rasmussen Foundation, and from the Danish National Research Foundation. LT and GJ thank the European commission for funding through the FP7 project SOLID, and the Swedish Research Council. FR acknowledges support from the Studienstiftung des deutschen Volkes.


\begin{thebibliography}{99}


\bibitem{ES} E. Schr\"{o}dinger,
\emph{Discussion of Probability Relations Between Separated Systems},
Naturwissenschaften \textbf{23}, 807 (1935).

\bibitem{NC} M. A. Nielsen and I. L. Chuang, \emph{Quantum computation and quantum information} (Cambridge University Press, Cambridge, 2000).



\bibitem{RevSC} R. J. Schoelkopf and S. M. Girvin, 
\emph{Wiring up quantum systems},
Nature \textbf{451}, 664 (2008).

\bibitem{Neeley} M. Neeley, R. C. Bialczak, M. Lenander, E. Lucero, M. Mariantoni, A. D. O'Connell, D. Sank, H. Wang, M. Weides, J. Wenner, Y. Yin, T. Yamamoto, A. N. Cleland, and J. M. Martinis, 
\emph{Generation of Three-Qubit Entangled States using Superconducting Phase Qubits}, 
Nature \textbf{467}, 570 (2010).

\bibitem{Fedorov} A. Fedorov, L. Steffen, M. Baur, M. P. da Silva, and A. Wallraff, 
\emph{Implementation of a Toffoli gate with superconducting circuits}, 
Nature \textbf{481}, 170 (2012).

\bibitem{Reed} M. D. Reed, L. DiCarlo, S. E. Nigg, L. Sun, L. Frunzio, S. M. Girvin, and R. J. Schoelkopf,
\emph{Realization of three-qubit quantum error correction with superconducting circuits}, 
Nature \textbf{482}, 382 (2012).

\bibitem{Houck} A. A. Houck, J. Koch, M. H. Devoret, S. M. Girvin, and R. J. Schoelkopf, 
\emph{Life after charge noise: recent results with transmon qubits}, 
Quantum Inf. Process. \textbf{8}, 105 (2009).

\bibitem{Chow} J. M. Chow, J. M. Gambetta, A. D. Corcoles, S. T. Merkel, J. A. Smolin, C. Rigetti, S. Poletto, G. A. Keefe, M. B. Rothwell, J. R. Rozen, M. B. Ketchen, and M. Steffen, 
\emph{Universal Quantum Gate Set Approaching Fault-Tolerant Thresholds with Superconducting Qubits}, 
Phys. Rev. Lett. \textbf{109}, 060501 (2012).

\bibitem{Rigetti} C. Rigetti, J. M. Gambetta, S. Poletto, B. L. T. Plourde, J. M. Chow, A. D. Corcoles, J. A. Smolin, S. T. Merkel, J. R. Rozen, G. A. Keefe, M. B. Rothwell, M. B. Ketchen, and M. Steffen,
\emph{Superconducting qubit in waveguide cavity with coherence time approaching 0.1 ms}, 
Phys. Rev. B \textbf{86}, 100506 (2012).

\bibitem{Poletto} S. Poletto, J. M. Gambetta, S. T. Merkel, J. A. Smolin, J. M. Chow, A. D. Córcoles, G. A. Keefe, M. B. Rothwell, J. R. Rozen, D. W. Abraham, C. Rigetti, and M. Steffen, 
\emph{Entanglement of Two Superconducting Qubits in a Waveguide Cavity via Monochromatic Two-Photon Excitation}, 
Phys. Rev. Lett. \textbf{109}, 240505 (2012).

\bibitem{Paik} H. Paik, D. I. Schuster, L. S. Bishop, G. Kirchmair, G. Catelani, A. P. Sears, B. R. Johnson, M. J. Reagor, L. Frunzio, L. I. Glazman, S. M. Girvin, M. H. Devoret, and R. J. Schoelkopf, 
\emph{Observation of High Coherence in Josephson Junction Qubits Measured in a Three-Dimensional Circuit QED Architecture}, 
Phys. Rev. Lett. \textbf{107}, 240501 (2011).

\bibitem{Sears} A. P. Sears, A. Petrenko, G. Catelani, L. Sun, H. Paik, G. Kirchmair, L. Frunzio, L. I. Glazman, S. M. Girvin, and R. J. Schoelkopf,
\emph{Photon shot noise dephasing in the strong-dispersive limit of circuit QED}, 
Phys. Rev. B \textbf{86}, 180504 (2012).



\bibitem{VWC} F. Verstraete, M. M. Wolf, and J. I. Cirac, 
\emph{Quantum computation and quantum-state engineering driven by dissipation}, 
Nature Phys. \textbf{5}, 633 (2009).

\bibitem{Kraus} B. Kraus, H. P. B\"{u}chler, S. Diehl, A. Kantian, A. Micheli, and P. Zoller, 
\emph{Preparation of entangled states by quantum Markov processes}, 
Phys. Rev. A \textbf{78}, 042307 (2008).

\bibitem{Diehl} S. Diehl, A. Micheli, A. Kantian, B. Kraus, H. P. B\"{u}chler, and P. Zoller, 
\emph{Quantum States and Phases in Driven Open Quantum Systems with Cold Atoms}, 
Nature Phys. \textbf{4}, 878 (2008).

\bibitem{Muller} M. M\"{u}ller, K. Hammerer, Y. L. Zhou, C. F. Roos, and P. Zoller, 
\emph{Simulating open quantum systems: from many-body interactions to stabilizer pumping}, 
New J. Phys. \textbf{13}, 085007 (2011).

\bibitem{DissMemory} F. Pastawski, L. Clemente, and J. I. Cirac, 
\emph{Quantum memories based on engineered dissipation}, 
Phys. Rev. A \textbf{83}, 012304 (2011).

\bibitem{DissRepeater} K.G.H. Vollbrecht, C. A. Muschik, and J. I. Cirac, 
\emph{Entanglement distillation by dissipation and continuous quantum repeaters}, 
Phys. Rev. Lett. \textbf{107}, 120502 (2011).



\bibitem{PHBK} M. B. Plenio, S. F. Huelga, A. Beige, and P. L. Knight, 
\emph{Cavity-loss-induced generation of entangled atoms}, 
Phys. Rev. A \textbf{59}, 2468 (1999). 

\bibitem{Clark} S. Clark, A. Peng, M. Gu, and S. Parkins, 
\emph{Unconditional Preparation of Entanglement between Atoms in Cascaded Optical Cavities}, 
Phys. Rev. Lett. \textbf{91}, 177901 (2003).

\bibitem{VB} G. Vacanti, and A. Beige, 
\emph{Cooling atoms into entangled states}, 
New. J. Phys. \textbf{11}, 083008 (2009).

\bibitem{WS} X. T. Wang, and S. G. Schirmer, 
\emph{Generating maximal entanglement between non-interacting atoms by collective decay and symmetry breaking}, 
arXiv:1005.2114 (2010).

\bibitem{KRS} M. J. Kastoryano, F. Reiter, and A. S. S\o rensen, 
\emph{Dissipative Preparation of Entanglement in Optical Cavities}, 
Phys. Rev. Lett. \textbf{106}, 090502 (2011).

\bibitem{Busch} J. Busch, S. De, S. S. Ivanov, B. T. Torosov, T. P. Spiller, and A. Beige, 
\emph{Cooling atom-cavity systems into entangled states}, 
Phys. Rev. A \textbf{84}, 022316 (2011).

\bibitem{RKS} F. Reiter, M. J. Kastoryano, and A. S. S\o rensen,
\emph{Driving two atoms in an optical cavity into an entangled steady state using engineered decay}, 
New J. Phys. \textbf{14}, 053022 (2012).



\bibitem{Parkins} A. S. Parkins, E. Solano, and J. I. Cirac, 
\emph{Unconditional Two-Mode Squeezing of Separated Atomic Ensembles}, 
Phys. Rev. Lett. \textbf{96}, 053602 (2006).

\bibitem{MPC} C. A. Muschik, E. S. Polzik, and J. I. Cirac, 
\emph{Dissipatively driven entanglement of two macroscopic atomic ensembles}, 
Phys. Rev. A \textbf{83}, 052312 (2011).

\bibitem{DallaTorre} E. G. Dalla Torre, J. Otterbach, E. Demler, V. Vuletic, and M. D. Lukin, 
\emph{Dissipative Preparation of Spin Squeezed Atomic Ensembles in a Steady State}, 
Phys. Rev. Lett. \textbf{110}, 120402 (2013).



\bibitem{PCZ} J. F. Poyatos, J. I. Cirac, and P. Zoller, 
\emph{Quantum Reservoir Engineering with Laser Cooled Trapped Ions}, 
Phys. Rev. Lett. \textbf{77}, 4728 (1996).

\bibitem{CBK} J. Cho, S. Bose, and M. S. Kim, 
\emph{Optical Pumping into Many-Body Entanglement}, 
Phys. Rev. Lett. \textbf{106}, 020504 (2011).

\bibitem{Barreiro} J. T. Barreiro, M. M\"{u}ller, P. Schindler, D. Nigg, T. Monz, M. Chwalla, M. Hennrich, C. F. Roos, P. Zoller, and R. Blatt, 
\emph{An open-system quantum simulator with trapped ions}, 
Nature \textbf{470}, 486 (2011).



\bibitem{AGZ} A. Gonzalez-Tudela, D. Mart\'{i}n-Cano, E. Moreno, L. Mart\'{i}n-Moreno, C. Tejedor, and F. J. Garc\'{i}a-Vidal, 
\emph{Entanglement of Two Qubits Mediated by One-Dimensional Plasmonic Waveguides}, 
Phys. Rev. Lett. \textbf{106}, 020501 (2011).

\bibitem{Gullans} M. Gullans, T. G. Tiecke, D. E. Chang, J. Feist, J. D. Thompson, J. I. Cirac, P. Zoller, and M. D. Lukin, 
\emph{Nanoplasmonic Lattices for Ultracold Atoms}, 
Phys. Rev. Lett. \textbf{109}, 235309 (2012).

\bibitem{GP} A. Gonzalez-Tudela, and D. Porras, 
\emph{Mesoscopic Entanglement Induced by Spontaneous Emission in Solid-State Quantum Optics},
Phys. Rev. Lett. 110, 080502 (2013) (2012).



\bibitem{Kiffner} M. Kiffner, U. Dorner, and D. Jaksch, 
\emph{Dissipative quantum-light-field engineering},
Phys. Rev. A \textbf{85}, 023812 (2012).


\bibitem{FossFeig} M. Foss-Feig, A. J. Daley, J. K. Thompson, and A. M. Rey,
\emph{Steady-state many-body entanglement of hot reactive fermions}, 
Phys. Rev. Lett. \textbf{109}, 230501 (2012).



\bibitem{Krauter} H. Krauter, C. A. Muschik, K. Jensen, W. Wasilewski, J. M. Petersen, J. I. Cirac, and E. S. Polzik,
\emph{Entanglement Generated by Dissipation and Steady State Entanglement of Two Macroscopic Objects},
Phys. Rev. Lett. \textbf{107}, 080503 (2011).



\bibitem{Zhang} J. Zhang, Y. Liu, C.-W. Li, T.-J. Tarn, and F. Nori, 
\emph{Generating stationary entangled states in superconducting qubits}, 
Phys. Rev. A \textbf{79}, 052308 (2009).

\bibitem{Li} P.-B. Li, S.-Y. Gao, and F.-L. Li, 
\emph{Engineering two-mode entangled states between two superconducting resonators by dissipation}, 
Phys. Rev. A \textbf{86}, 012318 (2012).

\bibitem{Xia} K. Xia, M. Macovei, and J. Evers, 
\emph{Stationary entanglement in strongly coupled qubits}, 
Phys. Rev. B \textbf{84}, 184510 (2011).

\bibitem{Murch} K. W. Murch, U. Vool, D. Zhou, S. J. Weber, S. M. Girvin, and I. Siddiqi, 
\emph{Cavity-assisted quantum bath engineering}, 
Phys. Rev. Lett. \textbf{109}, 183602 (2012).



\bibitem{Koch} J. Koch, T. M. Yu, J. Gambetta, A. A. Houck, D. I. Schuster, J. Majer, A. Blais, M. H. Devoret, S. M. Girvin, and R. J. Schoelkopf, 
\emph{Charge-insensitive qubit design derived from the Cooper pair box}, 
Phys. Rev. A \textbf{76}, 042319 (2007).

\bibitem{PhaseQubit} Y. Yin, Y. Chen, D. Sank, P. J. J. O`Malley, T. C. White, R. Barends, J. Kelly, E. Lucero, M. Mariantoni, A. Megrant, C. Neill, A. Vainsenchr, J. Wenner, A. N. Korotkov, A. N. Cleland, and J. M. Martinis,
\emph{Catch and release of microwave photon states}, 
Phys. Rev. Lett. \textbf{110}, 107001 (2013).

\bibitem{Kelly} W. Kelly, Z. Dutton, J. Schlafer, B. Mookerji, T. A. Ohki, J. S. Kline, and D. P. Pappas,
\emph{Direct Observation of Coherent Population Trapping in a Superconducting Artificial Atom}, 
Phys. Rev. Lett. \textbf{104}, 163601 (2010).


\bibitem{Groot} P. C. de Groot, J. Lisenfeld, R. N. Schouten, S. Ashhab, A. Lupascu, C. J. P. M. Harmans, and J. E. Mooij,
\emph{Selective darkening of degenerate transitions demonstrated with two superconducting quantum bits},
Nature Phys. \textbf{6}, 763 (2010).

\bibitem{Chow11}
J. M. Chow, A. D. Corcoles, J. M. Gambetta, C. Rigetti, B. R. Johnson, J. A. Smolin, J. R. Rozen, G. A. Keefe, M. B. Rothwell, M. B. Ketchen, and M. Steffen,
\emph{Simple All-Microwave Entangling Gate for Fixed-Frequency Superconducting Qubits},
Phys. Rev. Lett. \textbf{107}, 080502 (2011).

\bibitem{Filipp}
S. Filipp, A. F. van Loo, M. Baur, L. Steffen, and A. Wallraff,
\emph{Preparation of subradiant states using local qubit control in circuit QED},
Phys. Rev. A \textbf{84}, 061805 (2011).

\bibitem{EO} F. Reiter and A. S. S\o rensen,
\emph{Effective operator formalism for open quantum systems}, 
Phys. Rev. A \textbf{85}, 032111 (2012).



\bibitem{QuTIP} J. R. Johansson, P. D. Nation, and F. Nori,
\emph{QuTiP 2: A Python framework for the dynamics of open quantum systems},
Comput. Phys. Commun. \textbf{184}, 1234 (2013);
\emph{QuTiP: An open-source Python framework for the dynamics of open quantum systems},
\textbf{183}, 1760 (2012).



\bibitem{Leghtas} Z. Leghtas, U. Vool, S. Shankar, M. Hatridge, S. M. Girvin, M. H. Devoret, and M. Mirrahimi,
\emph{Stabilizing a Bell state of two superconducting qubits by dissipation engineering}, 
Phys. Rev. A \textbf{87}, 042315 (2013).


\end{thebibliography}
\end{document}